\documentclass{article}
\usepackage{graphicx} 
\usepackage{appendix}
\usepackage{amsmath}
\usepackage[sort&compress,numbers]{natbib}

\usepackage[hypertexnames=false]{hyperref}
\hypersetup{
    colorlinks=true,
    linkcolor=blue,
    filecolor=magenta,   
    citecolor=black,   
    urlcolor=cyan,
    }

\def\sin#1{\textrm{Sin}(#1)}
\def\cos#1{\textrm{Cos}(#1)}

\begin{document}
\def\thefootnote{\fnsymbol{footnote}}
\def\thetitle{Precise Measurement of $\delta$ at LBL Experiments Using Only The Neutrino Sector}
\def\autone{Emilio Ciuffoli}
\def\affa{Institute of Modern Physics, NanChangLu 509, Lanzhou 730000, China}

\begin{center}
{\large {\bf \thetitle}}

\bigskip

\bigskip


{\large \noindent  \autone{${}^{1}$}\footnote{emilio@impcas.ac.cn}}


\vskip.7cm

1) \affa\\

\end{center}

\begin{abstract}
    \noindent
Currently, the CP-violating phase $\delta$ is the least precisely known among the neutrino mixing parameters. In the coming years, accelerator neutrino experiments such as DUNE and Hyper-K will measure $\delta$, significantly increasing the precision. Such a measurement is usually performed by comparing the oscillation probabilities in the neutrino and antineutrino sectors, thereby directly observing the CP violation. This approach, however, has some downsides: in particular, it is more challenging to obtain good statistics in the antineutrino sector, leading to increased statistical errors. In principle, however, it is possible to determine $\delta$ by looking only at the neutrino sector, studying the energy dependence of the oscillation probability. We investigate this possibility in detail; we find that, if this approach is used, the main issue would be the degeneracies with other mixing parameters, which affect the sensitivity to $\delta$. Those parameters, however, have already been measured with great precision, which will increase even more in the next few years; if those constraints are taken into account by introducing Gaussian penalty terms in the $\chi^2$, it is possible to achieve better results focusing only on the neutrino sector, rather than using the standard approach. Those degeneracies could also be broken by exploring a wider range of $L/E$: this can be achieved, for instance, by increasing the neutrino energy. In this way, the total beam intensity could also increase due to the relativistic boost (however, the conversion efficiency could decrease, affecting the total luminosity). We find that if the relativistic boost increases the number of events in the high-energy configuration, the best performance would be achieved in such a set-up; otherwise, the optimal approach would be to focus only on the neutrino mode without changing the beam energy. 
\end{abstract}

\setcounter{footnote}{0}
\renewcommand{\thefootnote}{\arabic{footnote}}

\section{Introduction}

The neutrino mixing parameters have been accurately measured in recent years; such a precision will only increase in the next few years, with the results from JUNO (already operative)~\cite{JUNO:2015zny,JUNO:2022mxj}, Hyper-K~\cite{Hyper-Kamiokande:2018ofw} and DUNE~\cite{DUNE:2016hlj, DUNE:2015lol, DUNE:2016evb, DUNE:2016rla} (they will start data taking in the next few years). Among those parameters, the one that still has the largest uncertainty is the CP-violating phase $\delta$. The experiments most sensitive to such a parameter are accelerator neutrino experiments, or LBL (Long-Baseline) experiments. Currently, most of the information we have on $\delta$ comes from T2K~\cite{T2K:2023smv} and NO$\nu$A~\cite{NOvA:2025tmb}; the uncertainties are very large, however, and the best-fit value for $\delta$ depends strongly on the assumptions made on the neutrino mass ordering: if it is assumed to be normal, we have $\delta\simeq\pi$ (no CP violation), otherwise, $\delta\simeq3\pi/2$ (maximal CP violation), see, for example, Ref.~\cite{Esteban:2024eli}. In the next few years, the uncertainty on $\delta$ will decrease significantly, thanks to Hyper-K and DUNE~\cite{Hyper-Kamiokande:2025fci, DUNE:2026aaw, Agarwalla:2024kti}. An experiment with a similar set-up, using JUNO as a far detector and HIAF as a beam source, has also been proposed in China~\cite{An:2025lws}. In those experiments, the secondary particles created by the collision of the proton beam, in particular pions and kaons, are focused using magnetic horns and directed into a tunnel, where they decay in flight. The decays of $\pi^+$ and $K^+$ create $\nu_\mu$, while the decays of $\pi^-$ and $K^-$ yield $\bar{\nu}_\mu$. By changing the direction of the current in the horns, it is possible to select a beam constituted mostly of $\nu_\mu$ or $\bar{\nu}_\mu$ (plus some unavoidable beam contamination); these two regimes are called FHC and RHC, respectively, standing for Forward and Reverse Horn Current. $\delta$ is determined by comparing the oscillation probability $P_{\mu \rightarrow e}$ in the neutrino and antineutrino sector; in this way, we can observe directly the CP-violation caused by the mixing parameters, and it is easier to break eventual degeneracies. Such an approach, however, presents challenges as well; for instance, in the energy ranges used in LBL experiments, the cross section for $\bar{\nu}_e$ is $2-3$ times lower than the one for $\nu_e$, which means that the number of events observed in the antineutrino sector is be considerably lower: for this reason, for example, Hyper-K will spend 25\% (75\%) of the time in neutrino (antineutrino) mode~\cite{Hyper-Kamiokande:2025fci}. Moreover, other effects can create a difference between the oscillation probability in the neutrino and antineutrino sectors, such as
 the Mikheyev–Smirnov–Wolfenstein (MSW) effect, and it is not possible to distinguish between those spurious sources of CP violation and the one we would have if $\delta\neq0,\pi$.

In this paper, we investigate whether it is possible to obtain a precise measurement of $\delta$ at LBL experiments using only the neutrino sector. The main advantage of this set-up is that the number of events would increase significantly. On the other hand, the measurement would be more sensitive to the degeneracies with the other mixing parameters. The effect of those degeneracies can be mitigated by exploiting the synergies with other neutrino experiments. Indeed, those mixing parameters have already been measured with great precision~\cite{Esteban:2024eli}, which will continue to increase with the next generation of experiments (see, for example, Refs.~\cite{JUNO:2022mxj,Agarwalla:2024kti}). If these constraints are taken into account, we can improve the sensitivity to $\delta$ by  focusing only on the neutrino sector, rather than using the standard approach.

We also investigate if changing the energy of the neutrino beam, allowing us to probe different regions of the $L/E$ parameter space, could improve the sensitivity to $\delta$. Such an approach is somehow similar to the one used in $\mu$DAR (Muon Decay At rest) experiments, which were proposed a few years ago, see, for example, Refs.~\cite{Alonso:2010fs, LENA:2011ytb, Aberle:2013ssa, Ciuffoli:2014ika, Ciuffoli:2015uta}. Here $\bar{\nu}_\mu$, $\nu_\mu$ and $\nu_e$ are created via $\mu$DAR and $\pi$DAR (Pion Decay At Rest). However, since neutrinos are detected via Inverse Beta Decay, which is sensitive to $\bar{\nu}_e$, only the channel $\bar{\nu}_\mu\rightarrow\bar{\nu}_e$ can be observed. In such a set-up, the degeneracy with the other pull parameters is broken by measuring the oscillation probability at different baselines, sampling different ranges of $L/E$. In LBL experiments, it would not be easy (or cheap) to change the distance between the neutrino source and the far detector. In some instances, however, it may be possible to change the neutrino beam energy, thereby allowing us to obtain the same result. Indeed, using the magnetic horns, only a certain range of momenta is selected; by changing such a range, it would be possible to have a different energy spectrum for the neutrino beam. 

We will use some approximations to estimate the high-energy spectrum and the effect of the relativistic boost, rather than rely on accurate simulations. The purpose of this paper, indeed, is not to provide a precise estimation of the sensitivity to $\delta$ of a specific experiment, but rather to study, qualitatively, if and under which conditions the set-ups discussed could achieve better results. For an accurate estimation of the expected sensitivity of a specific experiment, however, more detailed studies would be required. 

Nonetheless, when possible, it is useful to consider realistic spectra, systematic errors, detector responses, etc., in our studies. For this purpose, we use GLoBES simulations~\cite{Huber:2004ka, Huber:2007ji}, using the AEDL files for the DUNE detector, provided in Ref.~\cite{DUNE:2021cuw}. There are several reasons why we choose this particular experiment. The DUNE far detector is placed on-axis on the NuMI beam~\cite{DUNE:2015lol}, and the possibility of changing the fluxes by varying the target and horn placement has already been discussed~\cite{Adamson:2015dkw}. Moreover, as we will see in Sec~\ref{sec::RelBoost}, it would be considerably more difficult to change the neutrino beam energy for off-axis detectors. Such a possibility has also already been considered in the literature in relation to other topics, for example to study Non Standard Interactions~\cite{Masud:2017bcf}, sterile neutrinos~\cite{Parveen:2023ixk}, and the possible synergies between DUNE and Hyper-K~\cite{Ballett:2016daj}. Finally, the neutrino energy spectrum when the detector is on-axis is considerably broader than its off-axis counterpart. This will allow DUNE to observe not only the first observation minimum but also the second, which would help break the degeneracies with the other mixing parameters. If those alternative set-ups are not preferable in such conditions, it is unlikely that they would work off-axis. On the other hand, if they can improve the sensitivity to $\delta$ (and, as we will see, they can), more detailed studies can be done to determine whether a particular experiment could benefit from such an approach.

This paper is structured as follows: in Sec.~\ref{sec::RelBoost}, we derive some useful formulas related to the energy of neutrinos produced via Decay in Flight (DIF), focusing in particular on how their energy and the luminosity of the beam depend on the energy of the primary particle and the off-axis angle of the detector. In Sec~\ref{sec::Theory}, we review some theoretical concepts, such as the oscillation probability for the appearance channel and the MSW effect. In Sec.~\ref{sec::Simulations}, we present the simulations set-up, explaining the different experimental configurations considered and in Sec.~\ref{sec::Sensitivity} we present our results. Finally, in Sec.~\ref{sec::Conclusions}, we summarize the conclusions.

\section{Relativistic Boost}\label{sec::RelBoost}

In this section, we derive some formulas for the spectrum of neutrinos produced via DIF, in particular, we focus on how the relativistic boost change the flux if the neutrinos are detected at an off-axis angle $\theta$. DIF neutrinos are usually created from the decay of pions and kaons, the exact ratio depends on the details of the accelerator. In this section, for simplicity, we consider only $\pi$ as the source particle.

Let us call $p^{\nu(\pi)}_{\mu}$ the four-momentum of the neutrino (pion) in the lab frame; let us also define the z-axis as the direction of the $\pi$:
\begin{equation}\label{Eq::defP}
 p^\nu_{\mu}=(E_\nu,E_\nu\sin{\theta}\sin{\phi},E_\nu\sin{\theta}\cos{\phi},E_\nu\cos{\theta}), \qquad p^\pi_{\mu}=(E_\pi,0,0,p_\pi),
\end{equation}
where $E_\pi^2=p_\pi^2+m_\pi^2$, $m_\pi$ being the pion mass, and the neutrino is assumed to be massless. 

\begin{figure}
    \centering
    \includegraphics[width=0.5\linewidth]{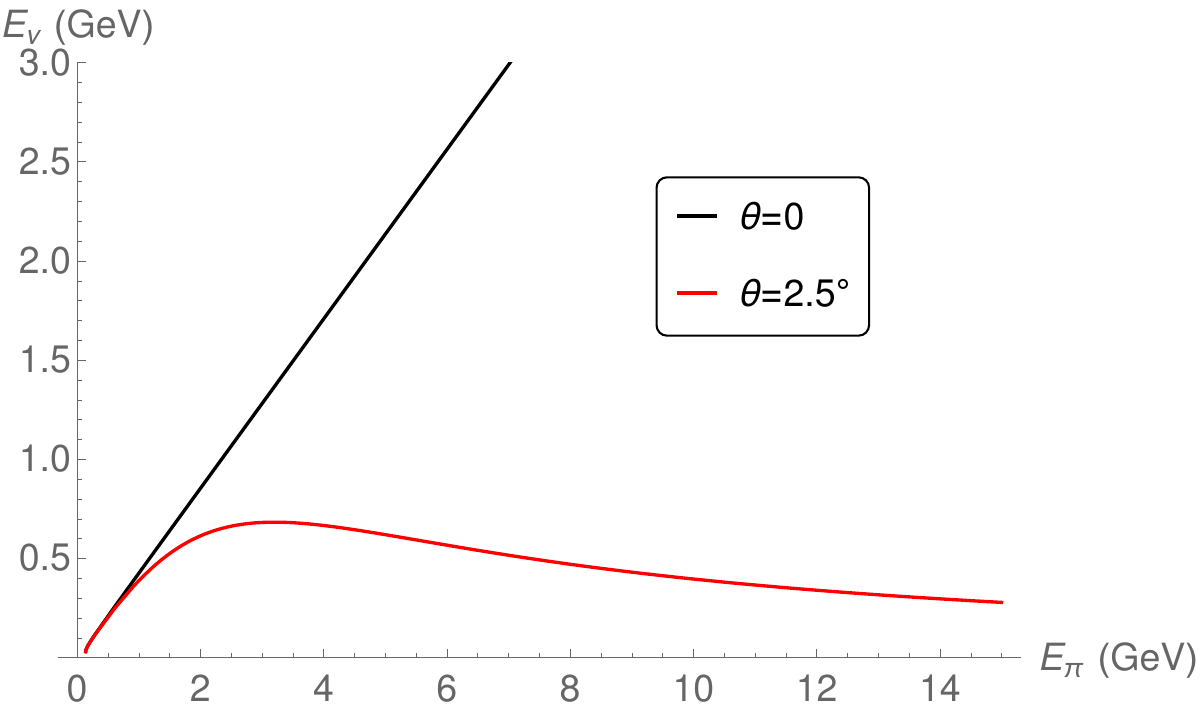}
    \caption{Neutrino energy $E_\nu$ as a function of pion energy $E_\pi$, for $\theta=0$ (detector on-axis) or $\theta=2.5^\circ$ (same off-axis angle as Hyper-K)}
    \label{fig::EnuVsEmu}
\end{figure}

Let us call $\tilde{p}^{\nu(\pi)}$ the neutrino (pion) momentum in the pion rest frame:
\begin{equation}\label{eq::DefPtilde}
 \tilde{p}^\nu_{\mu}=(\tilde{E}_\nu,\tilde{E}_\nu\sin{\tilde{\theta}}\sin{\phi},\tilde{E}_\nu\sin{\tilde{\theta}}\cos{\phi},\tilde{E}_\nu\cos{\tilde{\theta}}), \qquad \tilde{p}^\pi_{\mu}=(m_\pi,0,0,0).
\end{equation}
A Lorentz boost in the z-direction does not change the angle $\phi$. We can go from $\tilde{p}^X_\mu$ to $p^X_\mu$ ($X=\nu,\pi$) and vice versa via a Lorentz transformation, where $\beta$, $\gamma$ are\footnote{The expression given for $\beta$ corresponds to the transformation going from the lab frame to the pion rest frame.}
\begin{equation}
    \gamma= \frac{E_\pi}{m_\pi}, \qquad \beta=\frac{p_\pi}{E_\pi}.
\end{equation}
The neutrinos emitted in the pion decay are monoenergetic, with $\tilde{E}_\nu\sim 30$ MeV. In the rest frame of the laboratory, we have
\begin{equation}\label{eq::Enu}
    E_\nu=\frac{\tilde{E}_\nu}{\gamma(1-\beta\cos{\theta})}=\frac{\tilde{E}_\nu m_\pi}{E_\pi-p_\pi\cos{\theta}}.
\end{equation}
It is worth noting that if our far detector is off-axis, there is a maximum neutrino energy, as can be seen in Fig.~\ref{fig::EnuVsEmu}. The reason is that the higher the pion energy, the more collimated the neutrino beam. After a certain point, increasing the pion energy means that the more energetic neutrinos have directions very close to the beam direction and the energy of neutrinos traveling at an angle $\theta$ actually decreases.

Similarly, the angle $\tilde{\theta}$ can be expressed as a function of $\theta$ as
\begin{equation}
    \cos{\tilde{\theta}}=\frac{\cos{\theta}-\beta}{1-\beta\cos{\theta}}.
\end{equation}

We now obtain the angular distribution of neutrinos in the lab reference frame. In the pion rest frame,  such a distribution is trivial, since the neutrinos are created isotropically
\begin{equation}
    \frac{\textrm{d}N_\nu}{\textrm{d}\tilde{\Omega}}=\frac{N_\pi}{4\pi},
\end{equation}
where $N_\nu$ is the total number of $\pi$ decayed, and d$\tilde{\Omega}$ the infinitesimal solid angle in the pion rest frame. Let us assume that the pion beam is monoenergetic (the generalization to a non-trivial energy distribution is quite straightforward): in the laboratory rest frame, we have
\begin{equation}\label{eq::AngualDistritBoost}
    \frac{\textrm{d}N_\nu}{\textrm{d}{\Omega}}=\frac{\textrm{d}N_\nu}{\textrm{d}\tilde{\Omega}}\frac{\textrm{d}\cos{\tilde{\theta}}}{\textrm{d}\cos{\theta}}=
    \frac{N_\pi}{4\pi}\frac{E_\nu^2}{\tilde{E}_\nu^2}.
\end{equation}
This means that the relativistic boost changes the neutrino flux at a certain angle by a factor $E_\nu^2/\tilde{E}_\nu^2$. We should point out that $E_\nu$ has a non-trivial dependence on $\theta$, as illustrated in Eq.~(\ref{eq::Enu}), and, if we integrate over the solid angle, we would find $N_\nu=N_\pi$, as expected.

Finally, if the pion energy is shifted, the neutrino energy changes as well. We assume, for simplicity, that the detector is on-axis; our result, however, holds also if $\theta\neq 0$, as long as the neutrino energy is far away from the maximum ({\it i.e.} the linear regime is still valid, see Fig.~\ref{fig::EnuVsEmu}). We can also safely treat the pion as relativistic, since its energy is typically on the order of GeV. Under these assumptions, a shift $\Delta E_\pi$ in the pion energy produces a proportional shift $\Delta E_\nu$ in the neutrino energy, with
\begin{equation}
    \Delta E_\nu\sim\frac{2\tilde{E}_\nu}{m_\pi} \Delta E_\pi \sim 0.44 \Delta E_\pi.
\end{equation}

\section{Theoretical Overview}\label{sec::Theory}
Electron, muon, and tau neutrinos are not eigenstates of the free Hamiltonian, but they are created in a superposition of states
\begin{equation}
    \nu = 
\begin{pmatrix}
\nu_e \\
\nu_\mu \\
\nu_\tau
\end{pmatrix}
= U
\begin{pmatrix}
\nu_1 \\
\nu_2 \\
\nu_3
\end{pmatrix},
\end{equation}
where $\nu_\alpha$, $\alpha=e,\mu,\tau$ are the neutrinos in the interaction basis and $\nu_i$, $i=1,2,3$ are the neutrino mass eigenstates. Neutrinos oscillations are governed by seven parameters: three $\Delta m^2_{ij}=m_i^2-m_j^2$ ($i,j=1,2,3$), and four mixing angles, $\theta_{12}, \theta_{13}, \theta_{23}$ and $\delta$ that define the Pontecorvo–Maki–Nakagawa–Sakata (PMNS) mixing matrix  $U$ (or, simply, mixing matrix). In the usual parametrization,  $U$ is given by
\begin{eqnarray}\label{eq::PMNS}
    U &=&
\begin{pmatrix}
1 & 0 & 0 \\
0 & c_{23} & s_{23} \\
0 & -s_{23} & c_{23}
\end{pmatrix}
\begin{pmatrix}
c_{13} & 0 & s_{13}e^{i\delta} \\
0 & 1 & 0 \\
-s_{13}e^{-i\delta} & 0 & c_{13}
\end{pmatrix}
\begin{pmatrix}
c_{12} & s_{12} & 0 \\
-s_{12} & c_{12} & 0 \\
0 & 0 & 1
\end{pmatrix} \\
&=&
\begin{pmatrix}
c_{12}c_{13} & s_{12}c_{13} & s_{13}e^{-i\delta_{\text{CP}}} \\
-s_{12}c_{23} - c_{12}s_{13}s_{23}e^{i\delta_{\text{CP}}} & c_{12}c_{23} - s_{12}s_{13}s_{23}e^{i\delta_{\text{CP}}} & c_{13}s_{23} \\
s_{12}s_{23} - c_{12}s_{13}c_{23}e^{i\delta_{\text{CP}}} & -c_{12}s_{23} - s_{12}s_{13}c_{23}e^{i\delta_{\text{CP}}} & c_{13}c_{23}
\end{pmatrix},\nonumber
\end{eqnarray}
where $s_{ij}=\textrm{Sin}(\theta_{ij})$ and $c_{ij}=\textrm{Cos}(\theta_{ij}$). 

The oscillation probability in vacuum for neutrinos of energy $E$ that have traveled for a distance $L$ is given by
\begin{equation}\label{eq::DefProb}
    P_{\alpha \beta} = \delta_{\alpha \beta} 
- 4 \sum_{i < j}^{n} \operatorname{Re}[U_{\alpha i} U_{\beta i}^* U_{\alpha j}^* U_{\beta j}] \textrm{Sin}^2 \Delta_{ij}
+ 2 \sum_{i < j}^{n} \operatorname{Im}[U_{\alpha i} U_{\beta i}^* U_{\alpha j}^* U_{\beta j}] \textrm{Sin} 2\Delta_{ij},
\end{equation}
where $\Delta_{ij}=1.27\Delta  m_{ij}^2L/E$, with $\Delta m_{ij}^2$ expressed in eV$^2$, $L$ in km and $E$ in GeV (or, equivalently, $L$ in m and $E$ in MeV). If we want to consider the oscillation probability for antineutrinos, we would have to change $U\rightarrow U^*$. From Eq.~(\ref{eq::PMNS}), we can see that this is equivalent to the transformation $\delta\rightarrow -\delta$. This also means that CP violation ({\it i.e.} a difference in the oscillation probability of neutrinos and antineutrinos) is only possible in the appearance channel\footnote{This is only true for CP-violation induced by the mixing parameters. Other spurious effects, such as the matter effect, can lead to different oscillation probabilities for neutrinos and antineutrinos in the disappearance sector.}, namely when $\alpha\neq \beta$, as can be seen from Eq.~(\ref{eq::DefProb}) (different oscillation probabilities in the disappearance channel would require CPT violation, not just CP~\cite{Giunti:2010zs}). It should also be noted, however, that this does not mean that $\delta$ cannot appear at all in $P_{\alpha\rightarrow \alpha}$, only that such a quantity must be invariant if we change $\delta\rightarrow-\delta$. For instance, while $P_{e\rightarrow e}$ does not depend on $\delta$, $P_{\mu\rightarrow \mu}$ does contain terms proportional to Cos($\delta$)~\cite{Arafune:1997hd}; those, however, are subdominant, and the sensitivity to $\delta$ that could be achieved considering only this channel would be limited. In accelerator neutrino experiments, most of the sensitivity to $\delta$ comes from the appearance channel, which is what we are focusing on (the disappearance channel would still be considered in our simulations, however).

Eq.~(\ref{eq::PMNS}) defines the mixing matrix for vacuum oscillations. If neutrinos are created or propagate in a dense environment, however, the oscillation probability is modified by the MSW effect. The new mixing matrix is obtained by diagonalizing
\begin{equation}
H = \frac{1}{2E} U 
\begin{pmatrix} 
m_1^2 & &   \\ 
& m_2^2 &   \\ 
& & m_3^2  
\end{pmatrix}U^\dagger+
\begin{pmatrix} 
 V_e & &\\ 
 & 0 & \\ 
 & & 0 
\end{pmatrix},
\end{equation}
where $V_{e}=\sqrt{2}G_Fn_e$, $G_F$ is the Fermi constant and $n_e$ the electron density. The exact expression for the oscillation probability taking into account the MSW effect has been derived in~\cite{Zaglauer:1988gz}, however the formulas are quite complicated. For our purposes, it is sufficient to consider the expression for $P_{\mu\rightarrow e}$ obtained in~\cite{Arafune:1997hd} in the lowest-order approximation, assuming $\Delta_{21}\ll 1$, $aL/E\ll1$, where
\begin{equation}
    a=2\sqrt{2}G_Fn_e E= 7.6\times 10^{-5} \textrm{eV}^2 \frac{\rho}{1\textrm{gr/cm}^3}\frac{E}{1\textrm{GeV}}.
\end{equation}
Given these assumptions, we have\footnote{We have changed the notation to be consistent with the one used in our paper}
\begin{eqnarray}\label{eq::PmuE}
P(\mu \to e) &=& 4 \textrm{Sin}^2(\Delta_{31}) c_{13}^2 s_{13}^2 s_{23}^2 \left( 1 + \frac{a}{\Delta m_{31}^2} \cdot 2(1 - 2s_{13}^2) \right) \nonumber \\
&&+ 8\Delta_{21} \textrm{Sin} (\Delta_{31}) c_{13}^2 s_{13} c_{23} s_{23} c_{12} s_{12}\textrm{Cos} (\Delta_{31}+\delta)\nonumber\\
&& -4 \textrm{Sin}(2\Delta_{31}) c_{13}^2 s_{13}^2 s_{23}^2\Delta_{31}\frac{a}{\Delta m_{31}^2} (1 - 2s_{13}^2)\nonumber\\
&& -4\Delta_{21} \textrm{Sin} (2\Delta_{31}) c_{13}^2 s^2_{13} s^2_{23}s^2_{12}.
\end{eqnarray}
Eq.~(\ref{eq::PmuE}) gives us the oscillation probability for neutrinos; the one for antineutrinos can be obtained by changing the sign of $\delta$ and $a$.

In the second term of Eq.~(\ref{eq::PmuE}) there is a degeneracy between $\delta$ and $\Delta m_{31}^2$; a change of the latter, however, would affect the other terms as well, in particular the first one, which is dominant. Moreover, the neutrino energy spectrum at LBL experiments is centered around the 1-3 oscillation maximum, {\it i.e.} $\Delta_{31}\sim \pi/2$. As a consequence, if for instance we want to compensate for a change of $\delta$ of $10^\circ$, we would have to change $\Delta m_{31}^2$ of around 10\%, which would create significant tension with the current experimental data, since the precision on $\Delta m_{31}^2$ is of the order of 1\% (see, for example, Ref.~\cite{Esteban:2024eli}).

\section{Simulations Set-up}\label{sec::Simulations}

We used GLoBES to simulate the sensitivity of LBL experiments to $\delta$ in different scenarios. In particular, we used the AEDL files for the DUNE experiment, as described in Ref.~\cite{DUNE:2021cuw}. 
There, two separate files containing the signal and background spectra are provided, corresponding to the FHC and RHC modes. To test different configurations, we only change the RHC file, since we want to focus on the neutrino mode. 
In particular, we studied:
\begin{itemize}
    \item {\bf standard} configuration (which is sometimes denoted with the abbreviation ``std"), where we used the original AEDL files, without any modification;
    \item {\bf only neutrino} configuration: the spectra seen in the neutrino mode are now used both for FHC and RCH, but those spectra are not modified;
    \item {\bf high-energy} configuration: in this case in RHC we use the FHC spectra, but shifted toward higher energies by a factor $\Delta E$, {\it i.e.} $E\rightarrow E+\Delta E$. $\Delta E=0$, is equivalent to the ``only neutrino" case.
\end{itemize}
The first two cases are quite straightforward and can be simulated without modifying the fluxes in the AEDL files. The third case requires more assumptions, instead. Changing the magnetic horns, indeed, would introduce non-trivial modifications to the neutrino spectrum, which could be computed only with a detailed simulation of the beam production. For the purposes of this paper, however, we do not need a faithful representation of the spectra achievable in such a scenario. We approximate the ``high-energy" configuration by shifting the energy spectra to the right by a fixed amount $\Delta E$, without changing their shape. Another issue is the normalization of total flux for the ``high-energy" configuration. On one hand, as seen in Sec.~\ref{sec::RelBoost}, the relativistic boost focuses the neutrino beam, increasing its intensity. On the other hand, changing the set-up of the magnetic horns could, for example, reduce the conversion efficiency ({\it i.e.} the average number of neutrinos in the beam obtained per collision). For this reason, we consider two scenarios: keeping the total normalization unchanged, and multiplying the spectra by a global factor $(E_{peak}+\Delta E)^2/E_{peak}^2$, where $E_{peak}=2.5$ GeV is the peak energy of the unoscillated $\nu_\mu$ spectrum, as an estimation of the effect of the relativistic boost, see Eq.~(\ref{eq::AngualDistritBoost}). In this way, we are able to estimate how much such a boost could contribute to the sensitivity to $\delta$, and if it would be a crucial factor. It should also be noted that, even if the boost is not taken into account, the total number of events would still differ: on one hand, since we are farther away from the oscillation maximum, in the appearance channel the oscillation probability, on average, decreases; on the other hand, however, the cross-section increases. The expression ``with relativistic boost" is sometimes shortened to ``w/ RB" or simply ``RB", while the cases where it is not taken into account are indicated with ``NB" (No Boost). Finally, we note that in the high-energy configuration, we always assume that 50\% of the time is spent in FHC and 50\% is spent studying the high-energy spectrum. 

\begin{figure}
    \centering
    \includegraphics[width=0.4\linewidth]{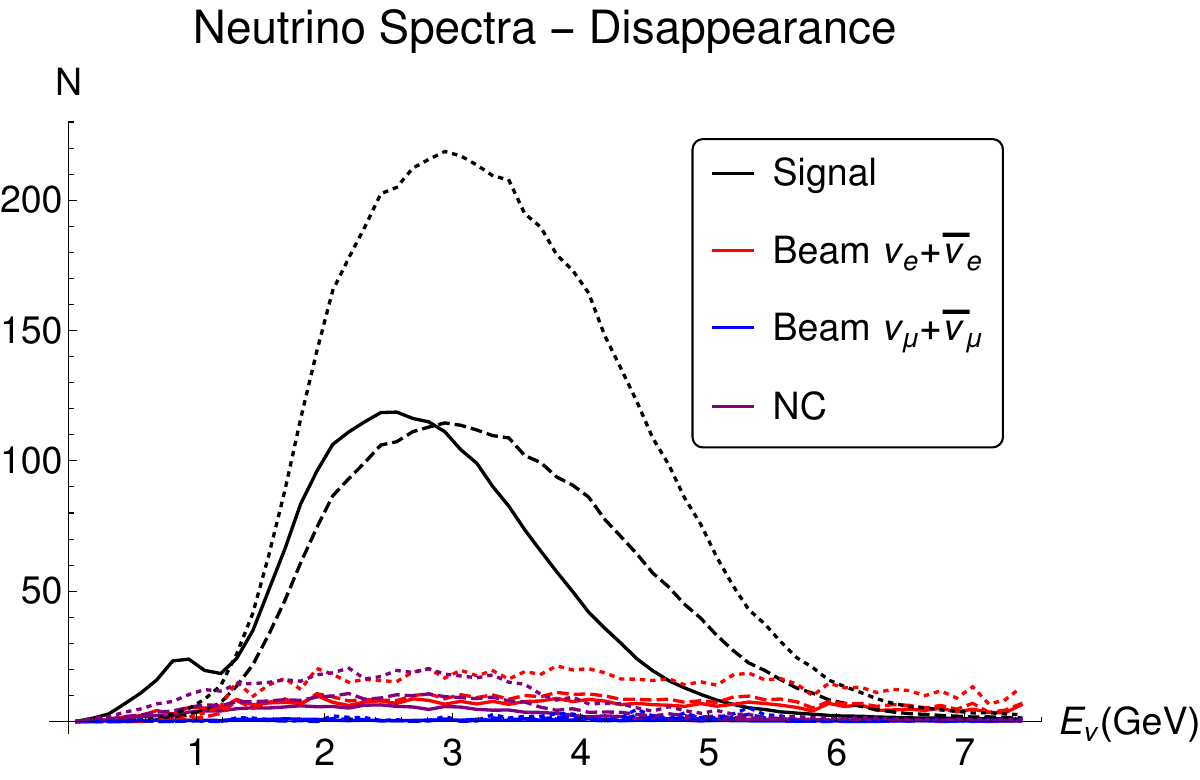}
    \includegraphics[width=0.4\linewidth]{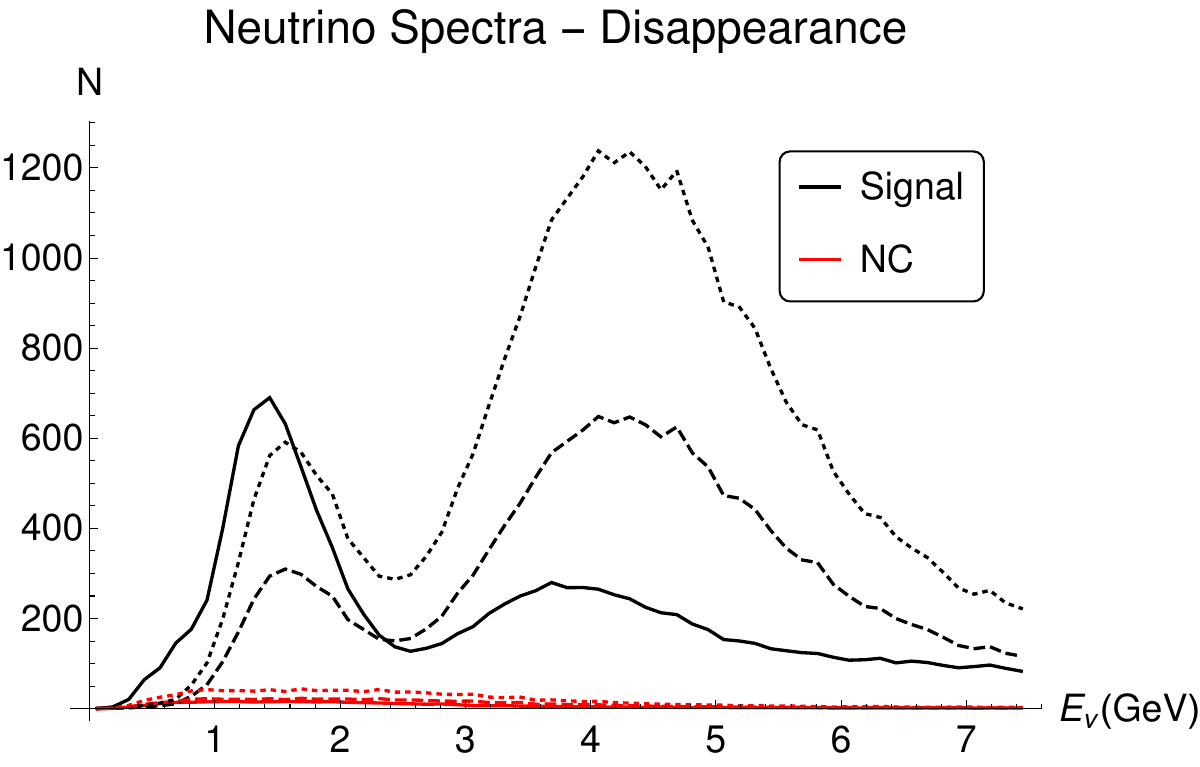}
    \caption{Expected number of signal and background events in the appearance (left panel) and disappearance (right panel) channels. The ``only neutrino" configuration is indicated by solid curves, the ``high-energy" configuration with (without) relativistic boost by dashed (dotted) curves. For the ``high energy" mode, only the spectra with $\Delta E=1$ GeV is reported.}
    \label{fig::Spectra}
\end{figure}

In our simulations, the expected spectrum was computed using the values of the mixing parameters (except $\delta$) reported in NuFit6.1~\cite{Esteban:2024eli}, which we present in Tab~\ref{tab::NuFit61}, along with the uncertainty for each parameter, defined as half of the dimension of the 1-$\sigma$ region. All the mixing angles (and relative uncertainties) are expressed in degrees, while the $\Delta m^2$'s are expressed in eV$^2$.
We have not made any assumptions on $\delta$, and the expected precision was computed for 36 different ``true" values of $\delta=\delta_{0}$, from 0 to 2$\pi$.  The neutrino mass ordering is assumed to be normal, since both global fits~\cite{Esteban:2024eli} and, more recently, the first results from JUNO~\cite{Yifang:Neutrino} indicate a preference for this scenario.  
\begin{table}[]
    \centering
    \begin{tabular}{|c|c|c|c|c|}
         \hline 
         $\theta_{12}$& $\theta_{13}$& $\theta_{23}$ & $\Delta m_{21}^2$ & $\Delta m_{31}^2$  \\\hline
         33.8$\pm$ 0.4 &8.62$\pm$ 0.11 &43.3$\pm$ 0.9 &(7.54$\pm$ 0.10)$\times10^{-5}$&(2.511$\pm$ 0.022)$\times10^{-3}$ \\ \hline
    \end{tabular}
    \caption{Values of the mixing parameters and relative uncertainties used in the simulations, from NuFit6.1~\cite{Esteban:2024eli}. The mixing angles are expressed in degrees, $\Delta m_{ij}^2$ in eV$^2$.}
    \label{tab::NuFit61}
\end{table}
The spectra for signal and background are reported in Fig.~\ref{fig::Spectra}, both with and without the relativistic boost, for the appearance (left panel) and the disappearance (right panel) channels, assuming $\delta=-\pi/2$.
\section{Sensitivity}\label{sec::Sensitivity}
In GLoBES, for each experiment, there are several ``rules" corresponding to each channel; in our case, we have the appearance and disappearance channels for both FHC and RHC modes. In our simulations, we use the Asimov data set~\cite{Cowan:2010js}: namely, we consider only the expected number of events, without taking into account statistical fluctuations. The formula used to compute $\chi^2$ is~\cite{Huber:2007ji}:
\begin{equation}\label{eq::defChi^2}
\chi^2(\delta)=\textrm{min}_{\vec{\alpha}}\sum_r\chi^2(O_r^i(\delta_{0},\vec{\alpha}_0),T_r^i(\delta,\vec{\alpha}))+\sum_j\frac{(\alpha_j-\alpha_{0,j})^2}{\sigma_{\alpha_j}^2}.
\end{equation}
where $\chi^2(O_r^i,T_r^i)$ is the standard Poisson $\chi^2$, $O_r^i(T_r^i)$ indicates the expected (fitted) number of events, and $r$ ($i$) is the rule (bin) number. Both $O_r^i$ and $T_r^i$ depend on $\delta$ and on some additional pull parameters, which are collected in the vector $\vec{\alpha}=\{\alpha_j\}$; in addition to the other mixing parameters, those include the matter density, the total flux normalization, etc..., see Ref.~\cite{DUNE:2021cuw} for details. $\vec{\alpha}_0$ and $\delta_{0}$ are the ``true" values of those parameters, {\it i.e.} the ones used to compute $O_r^i$, $\chi^2(\delta)$ is minimized over $\vec{\alpha}$, which, along with $\delta$, is used to compute $T_r^i$. Finally, the last term represents the Gaussian penalty term for the pull parameters, with $\alpha_{0,j}$ and $\sigma_{\alpha_j}$ being the central value and the standard deviation of the prior on the $j$-th component of $\vec{\alpha}$, respectively. We always assume that the central values are equal to the ``true" values of those parameters. For the mixing parameters, $\sigma_{\alpha_j}$ is equal to the uncertainty on $\alpha_j$ reported in Tab.~\ref{tab::NuFit61}, for the matter density, we assume an error of 2\%, the same value used in Ref.~\cite{DUNE:2021cuw}; and for the other pull parameters, such as the total flux normalization, we use the uncertainties reported in the AEDL files.

Since the Asimov data set is used, $\chi^2(\delta_{0})=0$ and the $\Delta\chi^2(\delta)$ is defined as
\begin{equation}
    \Delta\chi^2(\delta)=\chi^2(\delta).
\end{equation}
The sensitivity to $\delta$ is expressed as the expected value of the standard deviation $\sigma$, defined as
\begin{equation}\label{eq::DefSigma}
   \sigma=\frac{\sigma_++\sigma_-}{2}, \qquad \Delta\chi^2(\delta_{0}\pm\sigma_\pm)=1.
\end{equation}
The main advantage of this definition is that it is possible to summarize the sensitivity to $\delta$ in a single parameter. On the other hand, however, this is accurate only as long as we are in the Gaussian regime, which is often not the case. 
Despite some loss of information, however, such an approximation is required if we want to easily compare several different scenarios. This choice justify also the use of Gaussian priors in Eq.~(\ref{eq::defChi^2}): while this is not always the case (in particular for $\theta_{23}$), the deviations from Gaussianity, given our definition for $\sigma$, do not strongly affect our results.

\begin{figure}
    \centering
    \includegraphics[width=0.4\linewidth]{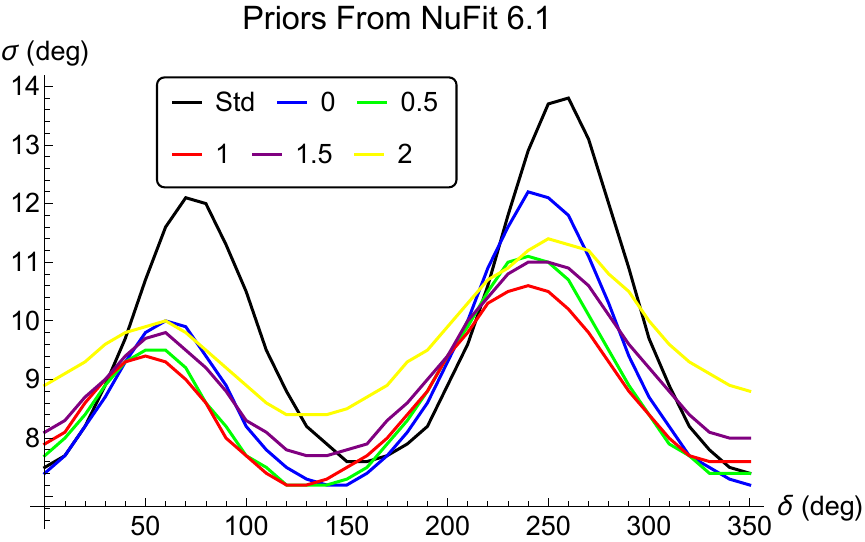}
    \includegraphics[width=0.4\linewidth]{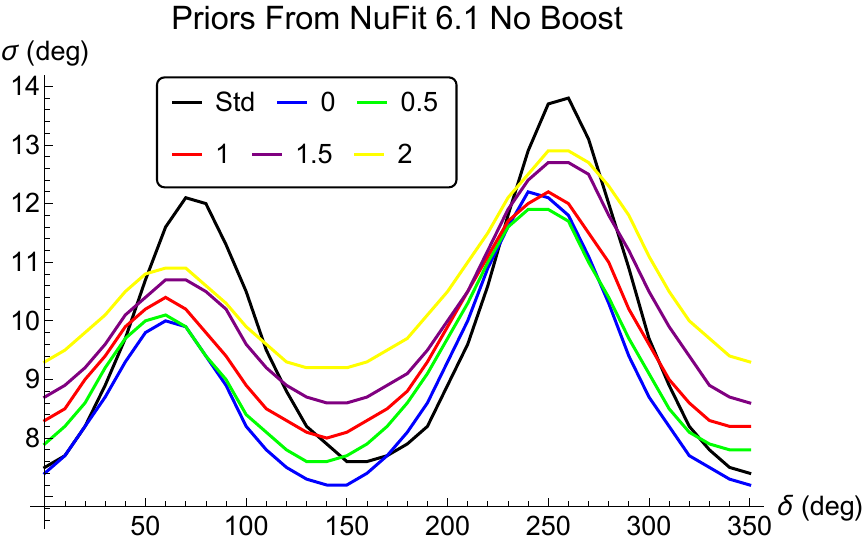}
    \caption{Sensitivity to $\delta$, expressed as the standard deviation $\sigma$ on the determination of its value (in degrees), considering different experimental configurations. The black curve corresponds to the standard configuration, while the others correspond to the ``high-energy" one, with different values of $\Delta E$ (expressed in GeV). $\Delta E=0$ corresponds to the ``only neutrino" configuration. In the left panel, the relativistic boost is taken into account, while in the right panel the total flux normalization is left unchanged.}
    \label{fig::SensitivityNuFit61}
\end{figure}

The sensitivity, computed using the configurations described above and the uncertainties on the mixing parameters reported in Tab.~\ref{tab::NuFit61}, is reported in Fig.~\ref{fig::SensitivityNuFit61}: in the left (right) panel, the relativistic boost was (not) taken into account. 
We can see that, if the present constraints on the mixing angles are taken into account, for most of the values of $\delta$ it is possible to achieve better results using the ``only neutrino" configuration, rather than with the standard set-up. If the unoscillated flux in the high-energy configuration is increased by the relativistic boost, this picture might change; in particular, under the assumptions we used, the best result is achieved by shifting the energy spectrum by $\Delta E = 1$ GeV.

\begin{figure}
    \centering
    \includegraphics[width=0.4\linewidth]{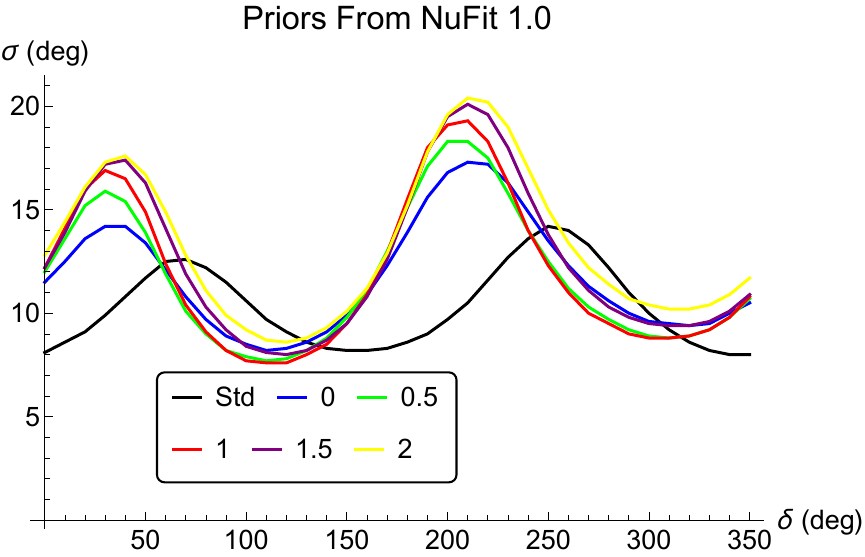}
    \includegraphics[width=0.4\linewidth]{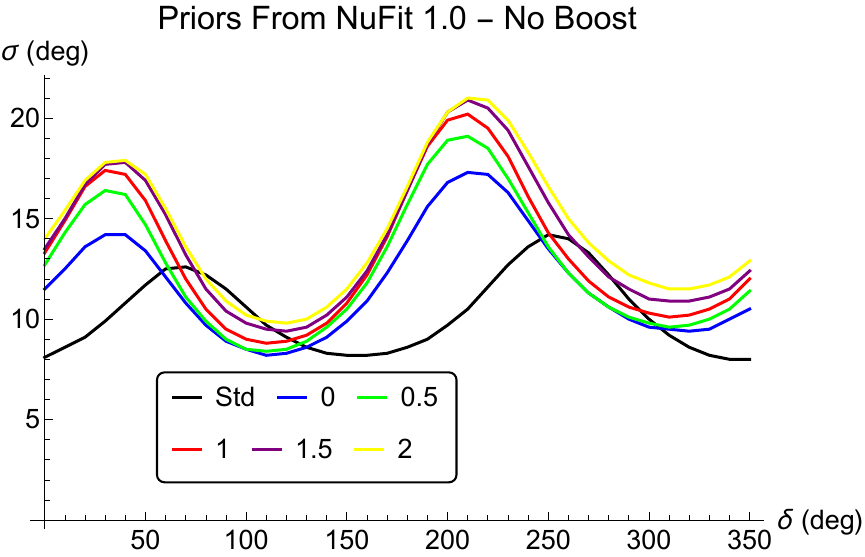}
    \caption{Same as Fig.~\ref{fig::SensitivityNuFit61}, but using the uncertainties from NuFit1.0}
    \label{fig::SensitivityNuFit10}
\end{figure}

\begin{figure}
    \centering
    \includegraphics[width=0.4\linewidth]{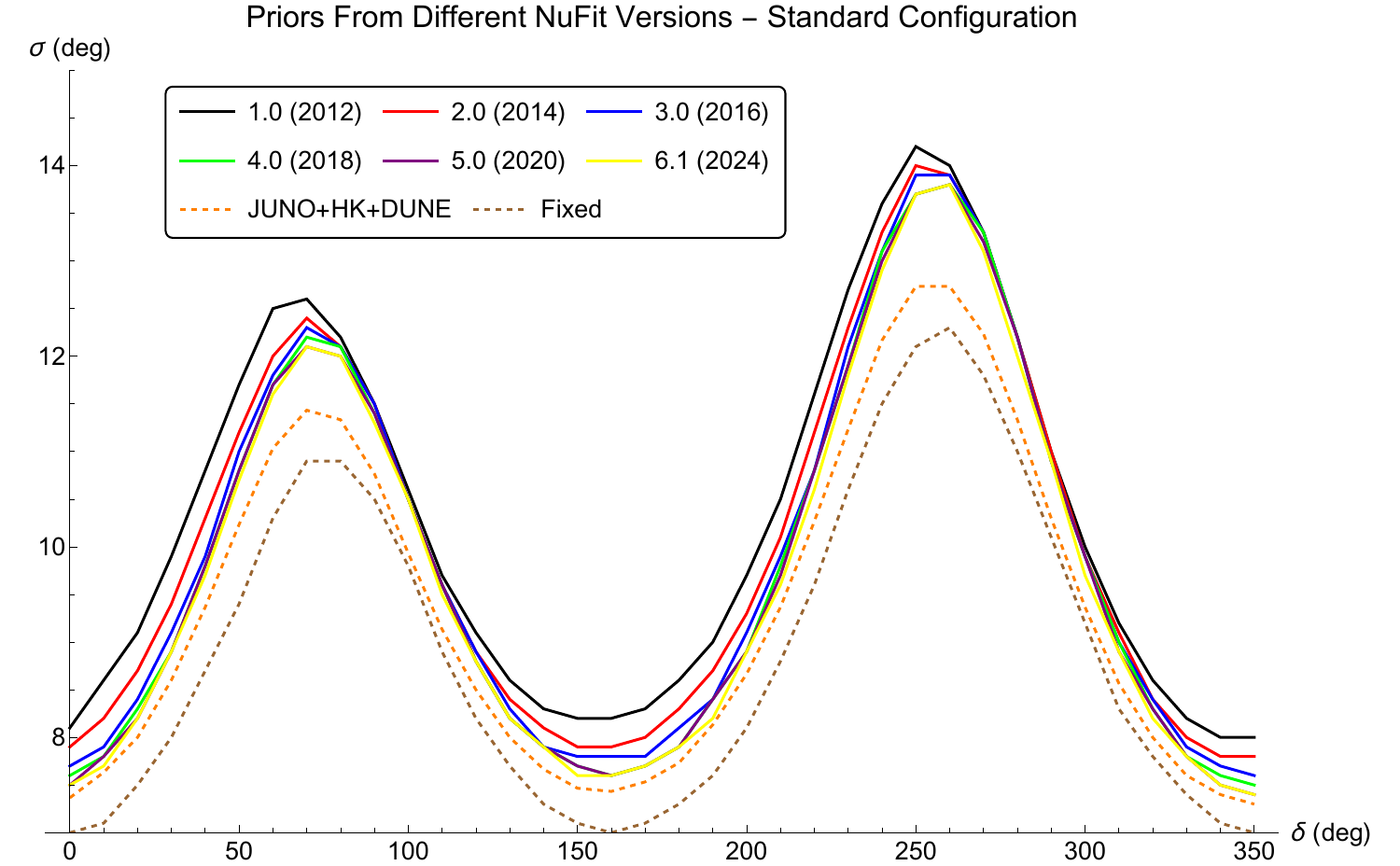}
    \includegraphics[width=0.4\linewidth]{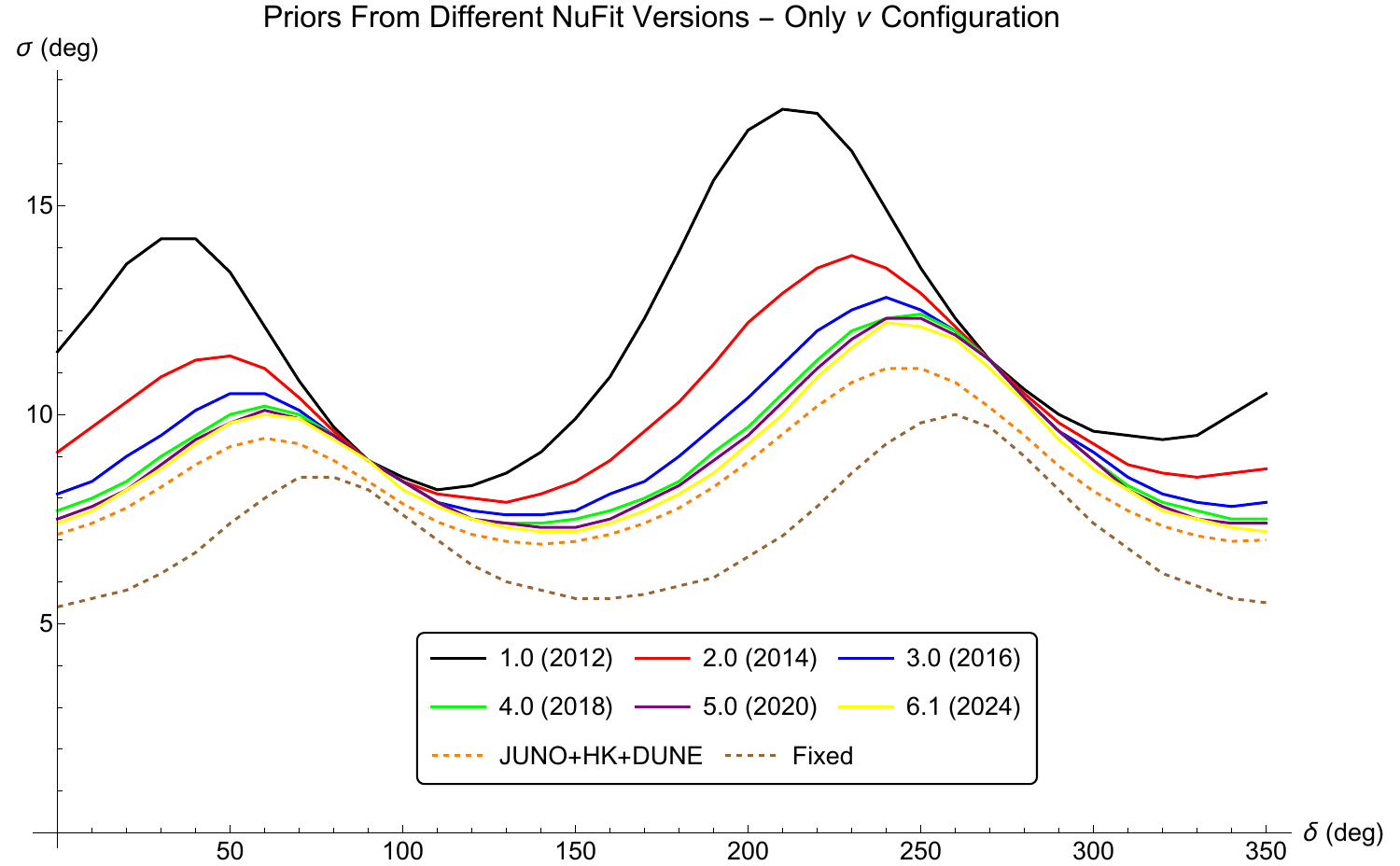}
    \caption{Left (right) panel: sensitivity to $\delta$ using the standard (``only neutrino") set-up. Different curves correspond to different versions of NuFit used. For comparison, we also included the results obtained considering the future JUNO+DUNE+Hyper-K results, as well as the case where all the mixing parameters are not minimized over ({\it i.e.} infinite precision).}
    \label{fig::SensitivityNorm}
\end{figure}

\begin{figure}
    \centering
    \includegraphics[width=0.4\linewidth]{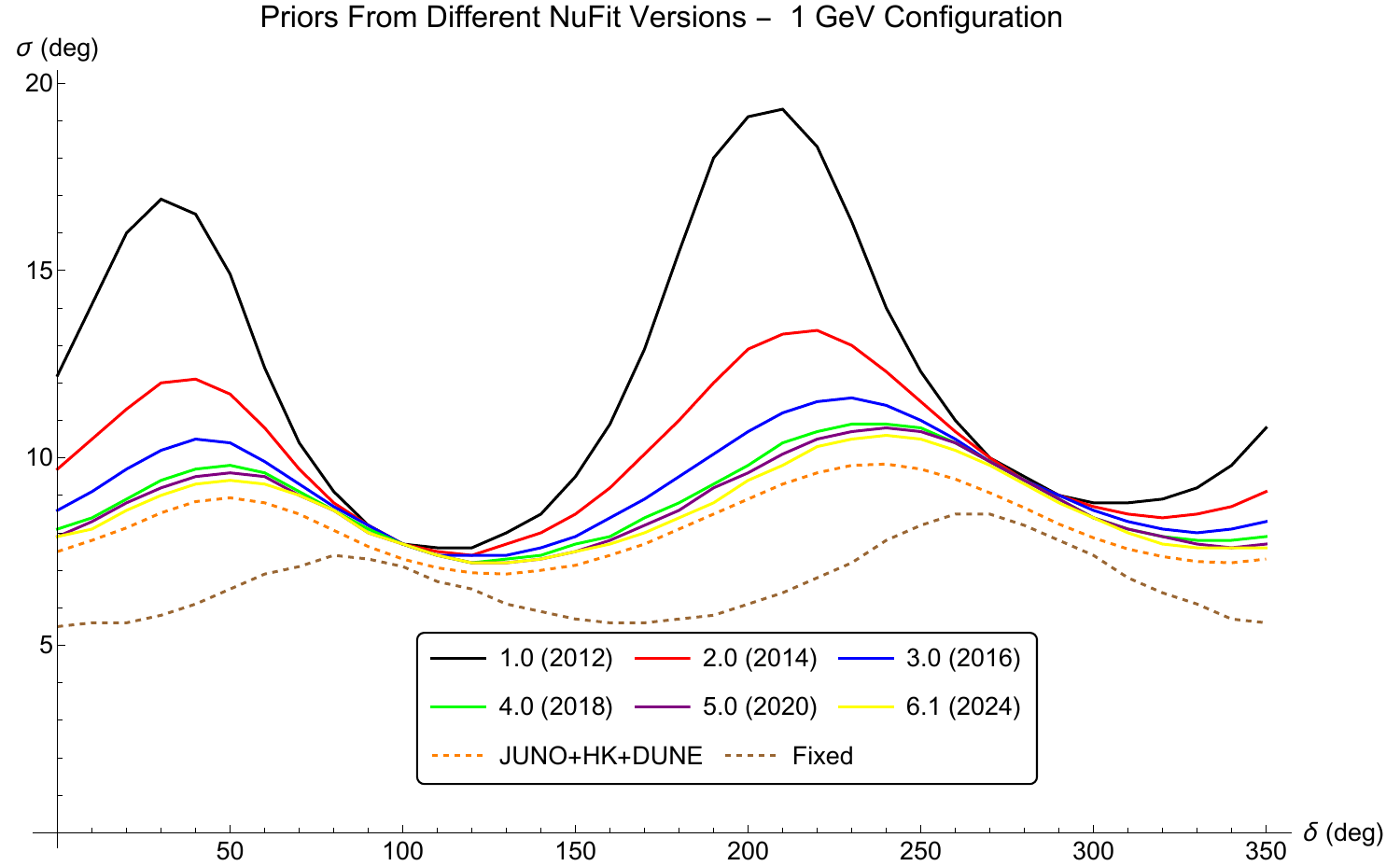}
    \includegraphics[width=0.4\linewidth]{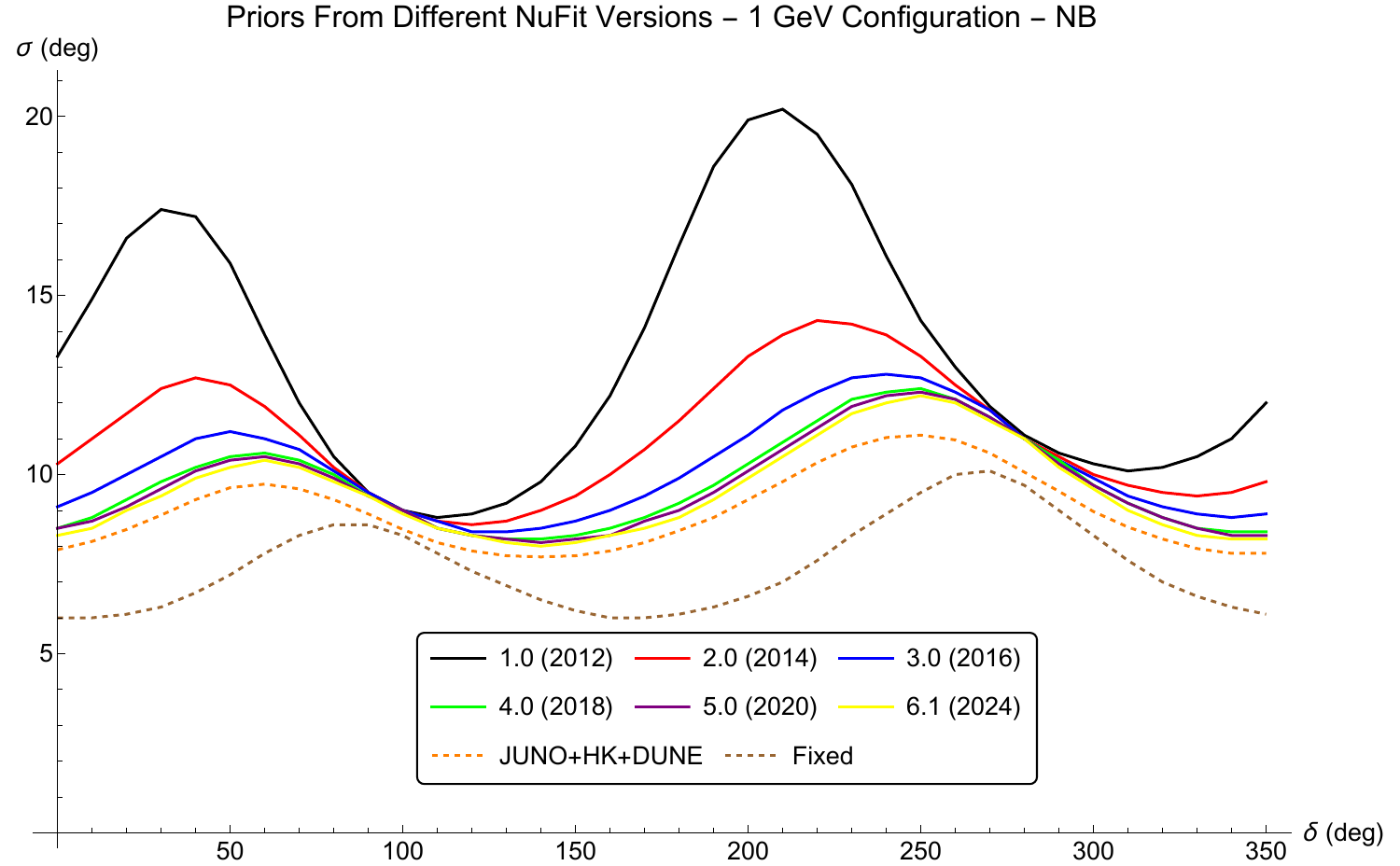}
    \caption{Same as Fig.~\ref{fig::SensitivityNorm}, but using the ``high-energy" mode with $\Delta E=1$ GeV. Left (right) panel: with(out) relativistic boost.}
    \label{fig::SensitivityOnlyNu}
\end{figure}

To study the effect of uncertainties on the other mixing parameters, we compute the expected sensitivity to $\delta$ using the uncertainties reported in previous versions of NuFit, ranging from 1.0 (June 2012) to 5.0 (July 2020), see Refs.~\cite{Gonzalez-Garcia:2012hef, Gonzalez-Garcia:2014bfa, Esteban:2016qun, Esteban:2018azc, Esteban:2020cvm}. We also take into account what precision can be achieved in the near future. Indeed, in the next few years, JUNO will significantly improve the precision on $\theta_{12}$, $\Delta m_{21}^2$, and $\Delta m_{31}^2$~\cite{JUNO:2022mxj}; moreover, the combined results of Hyper-K and DUNE will improve the precision on $\theta_{23}$ as well~\cite{Agarwalla:2024kti}. For convenience, we report those values in Tab.~\ref{tab::Uncert}.
\begin{table}[]
    \centering
    \begin{tabular}{|c|c|c|c|c|c|}
         \hline 
         source& $\sigma(\theta_{12})$& $\sigma(\theta_{13})$& $\sigma(\theta_{23})$ & $\sigma(\Delta m_{21}^2)$ & $\sigma(\Delta m_{31}^2)$  \\\hline
         NuFit1.0& 0.8 & 0.45 & 1.8 & $1.9\times10^{-6}$ & 6.9$\times 10^{-5}$ \\ \hline
         NuFit2.0& 0.77 & 0.21 &  2.3 & $1.8\times10^{-6}$ & 4.7$\times 10^{-5}$ \\ \hline
         NuFit3.0& 0.76 & 0.15 &  1.4 & $1.8\times10^{-6}$ & 4.0$\times 10^{-5}$ \\ \hline
         NuFit4.0& 0.77 & 0.13 &  1.0 & $2.1\times10^{-6}$ & 3.2$\times 10^{-5}$ \\ \hline
         NuFit5.0& 0.76 & 0.12 &  1.1 & $2.1\times10^{-6}$ & 2.7$\times 10^{-5}$ \\ \hline
         NuFit6.1&0.4 &0.11 &0.9 &$1.0\times10^{-6}$&$2.2\times10^{-5}$ \\ \hline
         JUNO &   0.1 & -    & - & $2.4\times10^{-7}$&$4.7\times10^{-6}$ \\ \hline
         HK+DUNE &  - & -    & 0.23 & -&- \\ \hline
    \end{tabular}
    \caption{Values of the uncertainties on the mixing parameters, including those from past global fits (NuFit1.0 to NuFit5.0), the most updated one (NuFit6.1), as well as the expected improvements from JUNO (6 years) and Hyper-K+DUNE. The uncertainties on the mixing angles are expressed in degrees; the ones on $\Delta m_{ij}^2$ are expressed in eV$^2$. }
    \label{tab::Uncert}
\end{table}
For comparison, in Fig.~\ref{fig::SensitivityNuFit10} we reproduce the same results shown in Fig.~\ref{fig::SensitivityNuFit61}, but using the uncertainties from NuFit1.0: in such a scenario, the standard configuration can outperform the others for almost all values of $\delta$.
In Figs.~\ref{fig::SensitivityNorm} and \ref{fig::SensitivityOnlyNu}, we can see how the sensitivity to $\delta$ for a given set-up changes with the different versions of Nufit, using the standard (Fig.~\ref{fig::SensitivityNorm}, left panel), ``only neutrino" (Fig.~\ref{fig::SensitivityNorm}, right panel) and ``high-energy" configuration (Fig.~\ref{fig::SensitivityOnlyNu}, $\Delta E=1$ GeV, left and right panel corresponding to the case with and without relativistic boost, respectively). We can see that the sensitivity to $\delta$ does not depend strongly on the uncertainties on the mixing parameters if the standard configuration is used. We should point out, however, that if we assume infinite precision, the improvement on the sensitivity to $\delta$ is not negligible, even with the standard set-up. On the other hand, if the ``only neutrino" or ``high-energy" set-ups are used, the dependence on uncertainties on the other mixing parameters is considerably stronger.
In Tab.~\ref{tab:SummarySensitivity}, we summarize some information on the sensitivity to $\delta$, depending on the configuration used and on the constraints on the other mixing parameters. We consider 4 configurations (standard, only neutrino, and $\Delta E=1$ GeV, with and without relativistic boost), and 4 different sets of uncertainties, namely NuFit1.0, NuFit6.1, JUNO+DUNE+HK (which, for space constraints, is shortened to ``JUNO"), and assuming infinite precision. For each combination, we report the minimum, maximum and average value of $\sigma$. For a given set of uncertainties, we also report the differences between the standard configuration and the others; in particular, we reported the difference between the average values of $\sigma$ (both as the absolute difference and as a percentage of the average value), as well as the maximum difference. Finally, in the last column, we have the percentage of times in which, using a given configuration, it is possible to achieve a higher sensitivity with respect to the standard one, assuming the same constraints on the other mixing angles are applied.
\begin{table}[]
    \centering
\begin{tabular}{|c|c|c|c|c|c|c|c|c|}
\hline
\text{Conf.} & \text{Prec.} & \text{$\sigma$ Min.} & $\sigma$ Max. & $\langle\sigma\rangle$ & \text{Diff.} & \text{Diff. (\%)} & \text{Max Diff.} & \text{\% Impr.} \\\hline\hline
\text{Std.} & \text{NuFit1.0} & 8 & 14.2 & 10.3 & 0. & 0. & 0. & 0. \\\hline
\text{OnlyNu} & \text{NuFit1.0} & 8.2 & 17.3 & 11.8 & -3.1 & -21.8 & 2.6 & 36. \\\hline
\text{1GeV-NB} & \text{NuFit1.0} & 8.8 & 20.2 & 13.3 & -6. & -42.3 & 2. & 28. \\\hline
\text{1GeV-RB} & \text{NuFit1.0} & 7.6 & 19.3 & 12.0 & -5.1 & -35.9 & 3.2 & 42. \\\hline\hline
\text{Std.} & \text{NuFit6.1} & 7.4 & 13.8 & 9.7 & 0. & 0. & 0. & 0. \\\hline
\text{OnlyNu} & \text{NuFit6.1} & 7.2 & 12.2 & 8.9 & 1.6 & 9.2 & 2.6 & 78. \\\hline
\text{1GeV-NB} & \text{NuFit6.1} & 8 & 12.2 & 9.6 & 1.6 & 9.2 & 2.2 & 47. \\\hline
\text{1GeV-RB} & \text{NuFit6.1} & 7.2 & 10.6 & 8.6 & 3.2 & 18.5 & 3.4 & 67. \\\hline\hline
\text{Std.} & \text{JUNO} & 7.3 & 12.7 & 9.3 & 0. & 0. & 0. & 0. \\\hline
\text{OnlyNu} & \text{JUNO} & 6.9 & 11.1 & 8.4 & 1.6 & 8.1 & 2.4 & 89. \\\hline
\text{1GeV-NB} & \text{JUNO} & 7.7 & 11.1 & 9.0 & 1.6 & 8.1 & 2.0 & 53. \\\hline
\text{1GeV-RB} & \text{JUNO} & 6.9 & 9.8 & 8.2 & 2.9 & 14.4 & 3.2 & 78. \\\hline\hline
\text{Std.} & \text{Inf. Prec.} & 7 & 12.3 & 8.9 & 0. & 0. & 0. & 0. \\\hline
\text{OnlyNu} & \text{Inf. Prec.} & 5.4 & 10 & 7.1 & 2.3 & 11.9 & 2.4 & 100. \\\hline
\text{1GeV-NB} & \text{Inf. Prec.} & 6 & 10.1 & 7.4 & 2.2 & 11.4 & 2.6 & 100. \\\hline
\text{1GeV-RB} & \text{Inf. Prec.} & 5.5 & 8.5 & 6.6 & 3.8 & 19.7 & 3.9 & 100. \\\hline
\end{tabular}
    \caption{Summary of the sensitivity to $\delta$. For a given configuration (standard, only neutrino, or $\Delta E=1$ GeV, with or without the relativistic boost, which were indicated with RB and NB, respectively) and a set of uncertainties on the mixing parameters (NuFit1.0, NuFit6.1, JUNO+DUNE+HK, which was shortened to ``JUNO", and assuming infinite precision), we have reported the minimum and maximum value of $\sigma$ over all the values of $\delta$ considered, as well as the average $\langle\sigma\rangle$. For each set of uncertainties, moreover, we have compared each configuration with the standard one, reporting the difference between $\langle\sigma\rangle$ (both as absolute value, and as a percentage), the maximum difference, and the fraction of times in which a certain configuration would outperform the standard set-up.}
    \label{tab:SummarySensitivity}
\end{table}
We can see that, focusing only on the neutrino sector and taking into account the current or future constraints on the other mixing parameters, the average sensitivity to $\delta$ would increase, with an improvement of the order of 1.5$^\circ -$3$^\circ$ (10-20\% better) in most of the cases (with $\Delta E=1$ GeV, w/ RB and assuming infinite precision, the improvement would be 3.8$^\circ$). The maximum difference with the standard configurations is around 2.4$^\circ$-2.6$^\circ$ for the ``only neutrino" configuration and around 3.2$^\circ$-3.4$^\circ$ for $\Delta E=1$ GeV, w/ RB. We should also note that, while the configuration with $\Delta E=1$ GeV w/ RB offers, in average, the best improvement on the sensitivity to $\delta$, the ``only neutrino" configuration is statistically more likely to yield better results: in other words, the ``high-energy" configuration can give a significantly increase the sensitivity for certain values of $\delta$, while the improvement obtained using the ``only neutrino" one is more consistent, even if lower.

\begin{figure}
    \centering
    \includegraphics[width=0.32\linewidth]{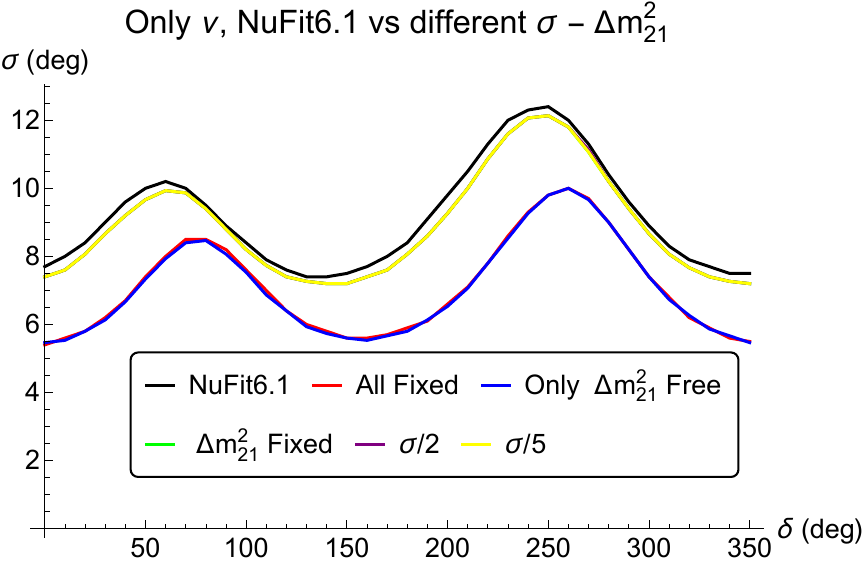}
    \includegraphics[width=0.32\linewidth]{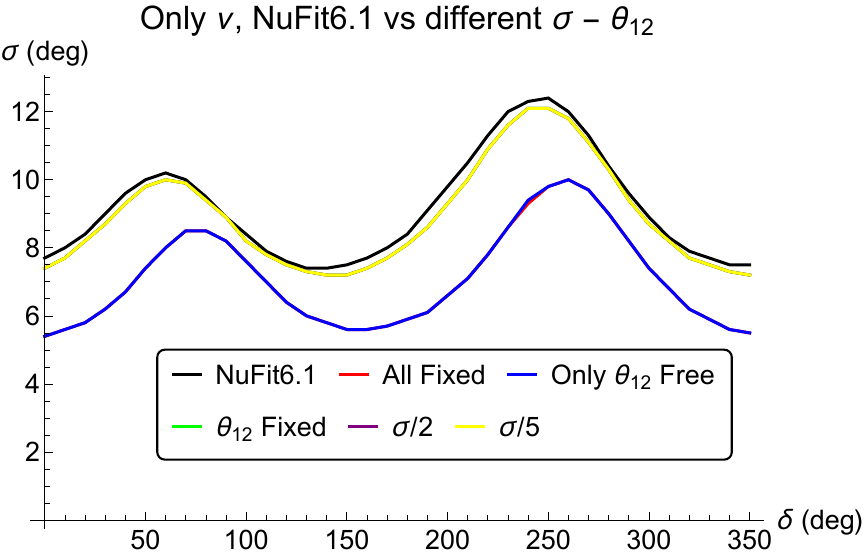}
    \includegraphics[width=0.32\linewidth]{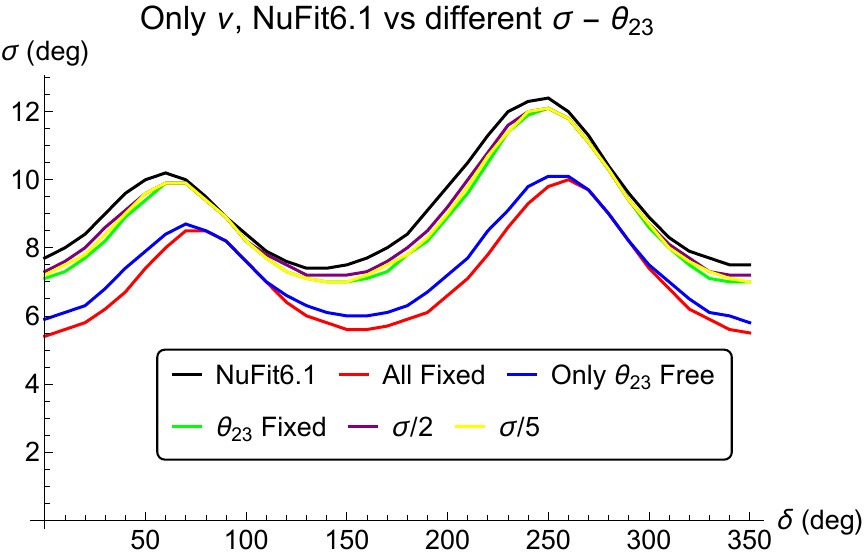}
    \caption{Sensitivity to $\delta$ changing the uncertainty on $\Delta m_{21}^2$ (left panel), $\theta_{12}$ (central panel) and $\theta_{23}$ (right panel), using the ``only neutrino" configuration.}
    \label{fig::12-23}
\end{figure}

\begin{figure}
    \centering
    \includegraphics[width=0.4\linewidth]{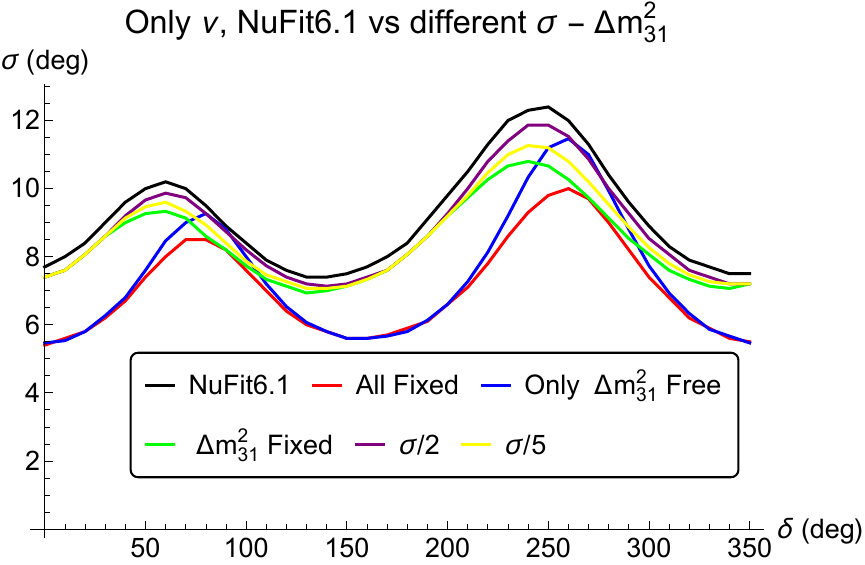}
    \includegraphics[width=0.4\linewidth]{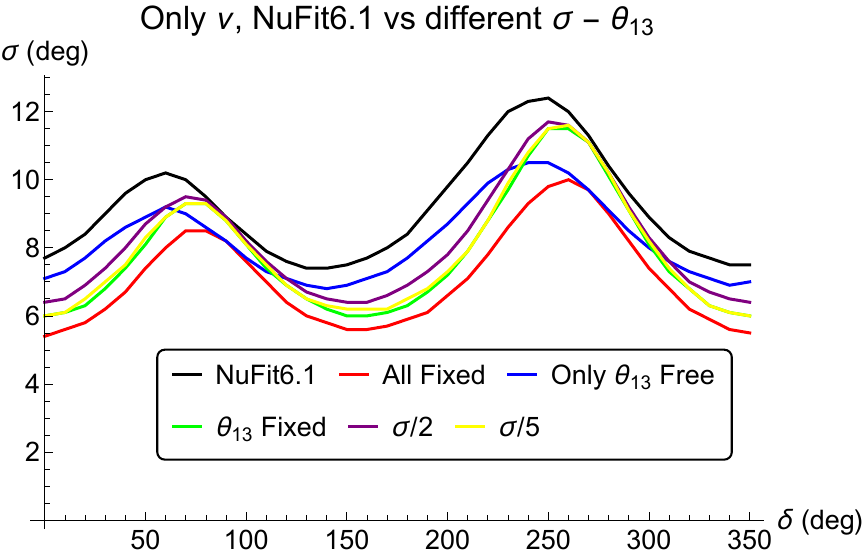}
    \caption{Same as Fig.~\ref{fig::12-23}, but considering the uncertainties on $\Delta m_{31}^2$ (left panel) and $\theta_{13}$ (right panel)}
    \label{fig::13}
\end{figure}

 We now study separately how the uncertainties on each mixing parameter can affect the measurement, using the ``only neutrino" configuration and considering several cases
\begin{enumerate}
    \item As a benchmark, we use the case where all the uncertainties on the mixing parameters are taken from NuFit6.1, corresponding to the blue curve in Fig.~\ref{fig::SensitivityNuFit61}.
    \item We divide the $\sigma$ used for a certain parameter by 2 or 5, while leaving the other $\sigma$'s unchanged. We also keep that parameter fixed (corresponding to infinite precision).
    \item We consider the case where all the parameters are fixed except one.
    \item We also report the case in which all the parameters are fixed (corresponding to the brown dashed curve in Fig~\ref{fig::SensitivityNorm}, right panel).
\end{enumerate}
Comparing 1) to 2), we can see how improving the precision of a given parameter can change the sensitivity to $\delta; $ comparing 3) to 4), we can deduce how much a given parameter alone can affect the measurement, if no degeneracies with other parameters are present. The results are reported in Fig.~\ref{fig::12-23} ($\Delta m_{21}^2$, $\theta_{12}$ and $\theta_{23}$ and Fig.~\ref{fig::13} ($\Delta m_{31}^2$ and $\theta_{13}$). 
We can see that the effects of the uncertainties on $\Delta m^2_{21}$ and $\theta_{12}$ are extremely marginal; this is to be expected, since the contribution of 1-2 oscillations to Eq.~(\ref{eq::PmuE}) are suppressed. The uncertainty on $\theta_{23}$ is not extremely relevant, either: the reason is that, as can be seen from Eq.~(\ref{eq::PmuE}), the term that depends on $\delta$ is proportional to $2c_{23}s_{23}=\textrm{Sin}(2\theta_{23})$. However, since $\theta_{23}\sim45^\circ$, a small shift in $\theta_{23}$ would not modify significantly the value of Sin($2\theta_{23}$). The octant degeneracy, on the other hand, could lead to a local minimum, as we will discuss later. The effect of the uncertainties on $\Delta m_{31}^2$ and $\theta_{13}$ is much larger, as we can see in Fig.~\ref{fig::13}. Indeed, as discussed before, there is a (partial) degeneracy between a shift of $\delta$ and a change of $\Delta m_{31}^2$, so it is understandable that the uncertainty on the latter can affect the precision of the former. The term of $P_{\mu\rightarrow e}$ that contains $\delta$, moreover, is proportional to $2c_{13}^2s_{13}=c_{13}\textrm{Sin}(2\theta_{13})$, which is much more susceptible to shifts of $\theta_{13}$. 

\begin{figure}
    \centering
    \includegraphics[width=\linewidth]{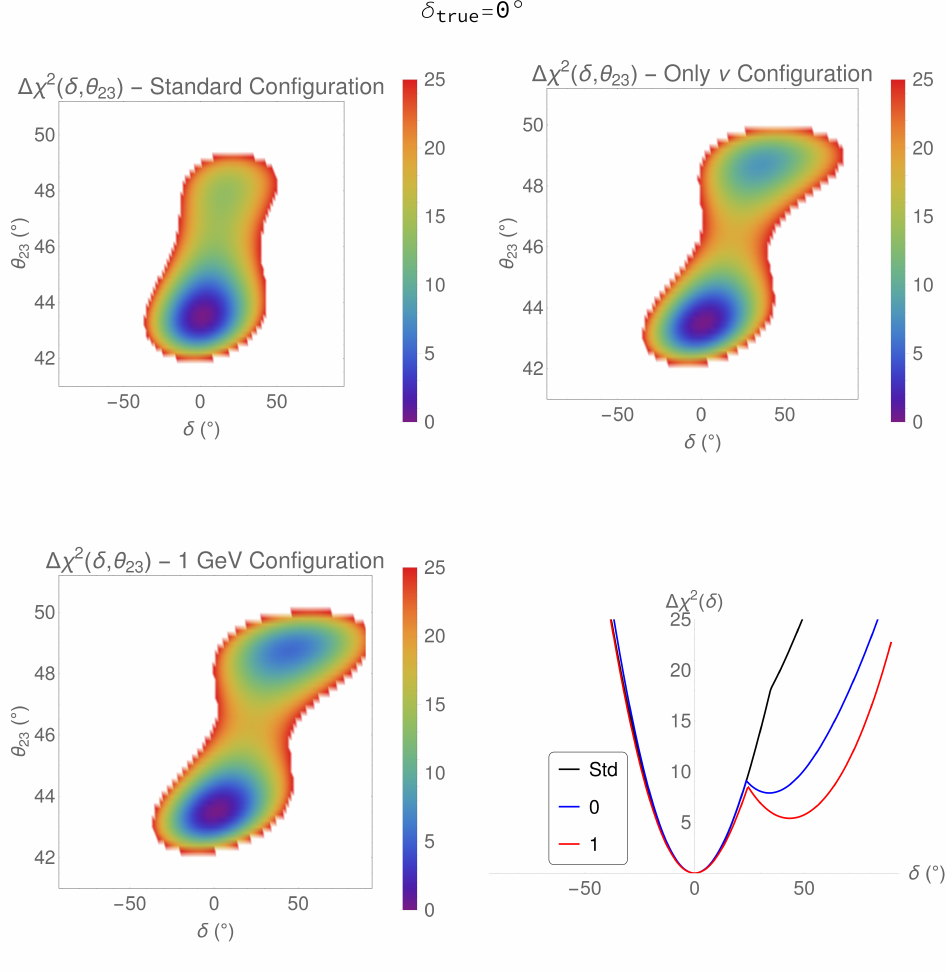}
    \caption{$\Delta\chi^2(\delta,\theta_{23})$ assuming $\delta_0=0^\circ$ and using the standard (top-left panel), only neutrino (top-right) or high energy (bottom-left) configuration. For the latter, we used $\Delta E=1 GeV$ and took into account the relativistic boost. No penalty term for $\theta_{23}$ were considered. Bottom-right panel: $\textrm{min}_{\theta_{23}}(\Delta\chi^2).  $}
    \label{fig::Density-1}
\end{figure}

\begin{figure}
    \centering
    \includegraphics[width=\linewidth]{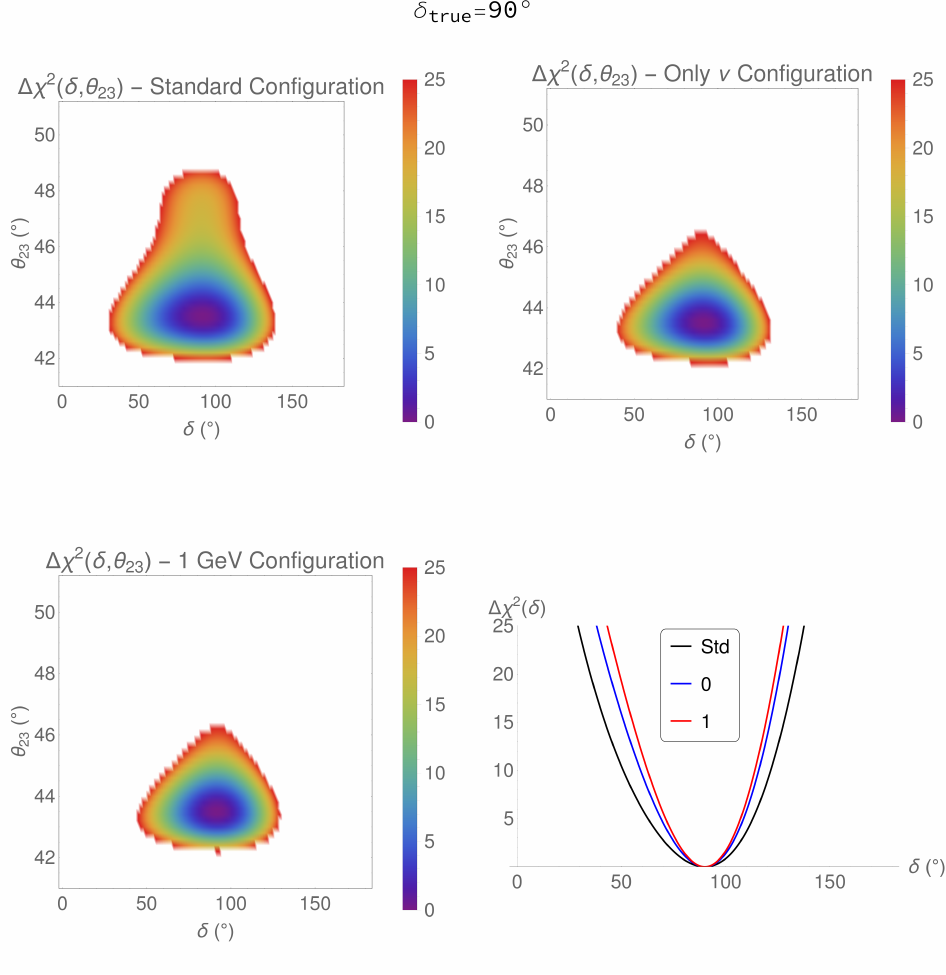}
    \caption{Same as Fig.~\ref{fig::Density-1}, but for $\delta_0=90^\circ$  }
    \label{fig::Density-2}
\end{figure}

\begin{figure}
    \centering
    \includegraphics[width=\linewidth]{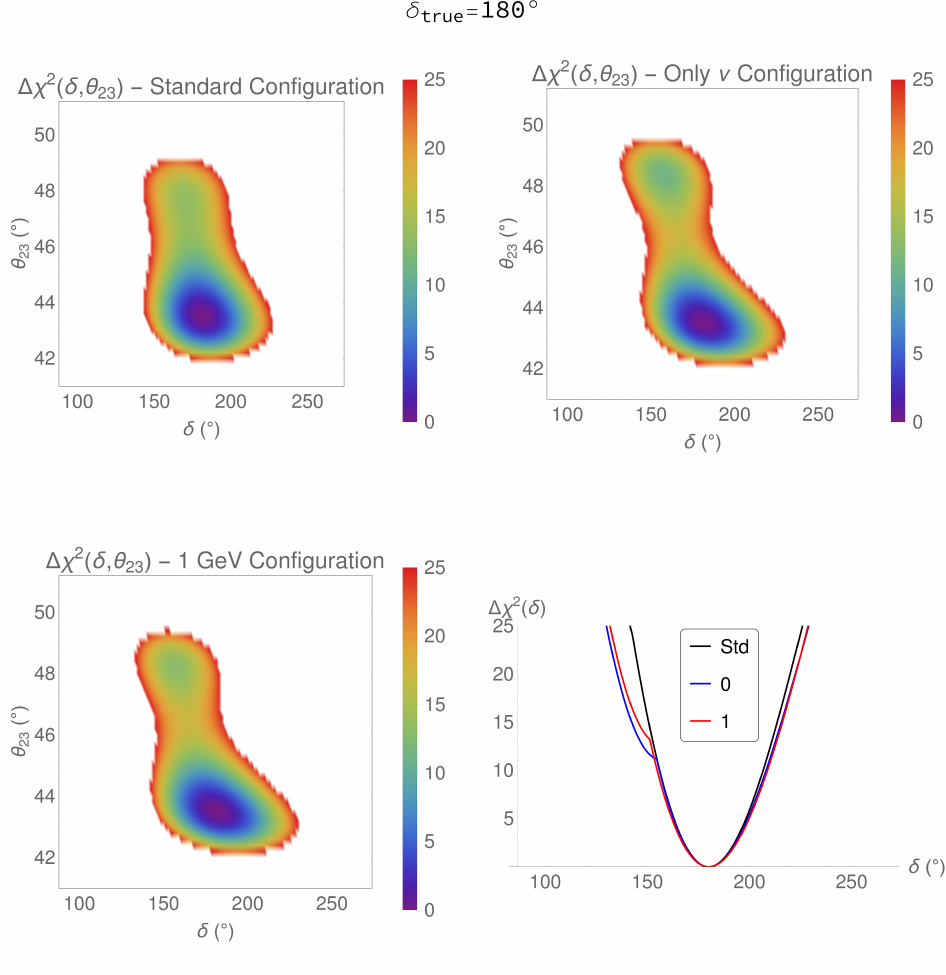}
    \caption{Same as Fig.~\ref{fig::Density-1}, but for $\delta_0=180^\circ$  }
    \label{fig::Density-3}
\end{figure}

\begin{figure}
    \centering
    \includegraphics[width=\linewidth]{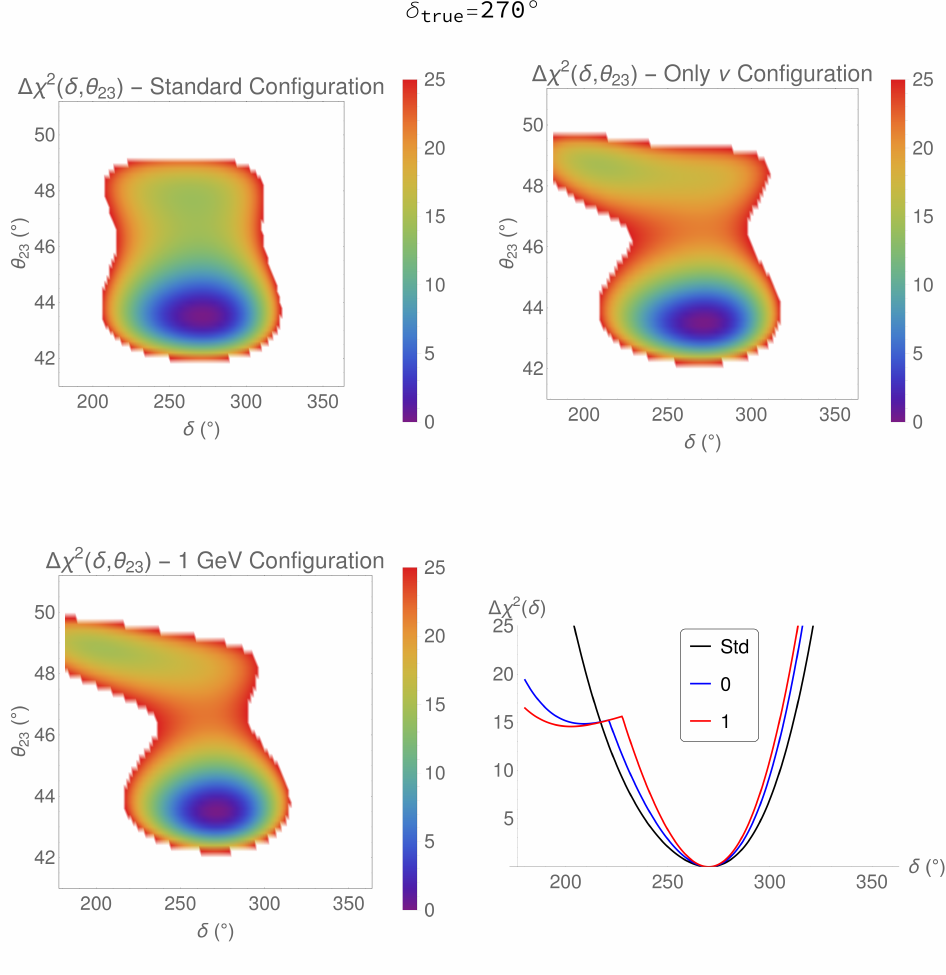}
    \caption{Same as Fig.~\ref{fig::Density-1}, but for $\delta_0=270^\circ$  }
    \label{fig::Density-4}
\end{figure}

Finally, we want to check the effect of the octant degeneracy. We compute $\Delta\chi^2(\delta,\theta_{23})$, {\it i.e.} without minimizing over $\theta_{23}$, using the uncertainties on the mixing parameters from NuFit6.1. In this case, however, no penalty term for $\theta_{23}$ is included in the computation of $\chi^2$. In Figs.~\ref{fig::Density-1}-\ref{fig::Density-4} we show the results for $\delta_{0}=0^\circ,90^\circ,180^\circ,270^\circ$; for each figure (corresponding to a specific value of $\delta_0$), in the bottom-right panel we have $\Delta\chi^2(\delta)=\textrm{min}_{\theta_{23}}\Delta\chi^2(\delta,\theta_{23})$; in the top-left, top-right and bottom-left panels, instead, we have the density plot of $\Delta\chi^2(\delta,\theta_{23})$ obtained with the standard, only neutrino and high energy configuration ($\Delta E=1$ GeV, RB), respectively. As expected, if we do not use the standard configuration, the effect of the degeneracy with the octant on the sensitivity to $\delta$ is larger. In particular, for $\delta_0=0^\circ, 270^\circ$, an additional minimum is present, ranging from $\Delta\chi^2\sim5.4$ (high-energy configuration, $\delta_0=0^\circ$) to $\Delta\chi^2\sim15$ (only neutrino configuration, $\delta_{0}=270^\circ$). Currently, even if we include a (non-Gaussian) prior on $\theta_{23}$ based on the global fit results in our $\chi^2$, it would not change this scenario significantly: indeed, while there is currently a weak preference for the lower octant, the statistical significance is quite low, {\it i.e.}  $\Delta\chi^2\sim1$~\cite{Esteban:2024eli}. Additional data, however, could substantially decrease the relevance of this minimum: for instance, both DUNE and Hyper-K are expected to exclude the wrong octant at around $ 3 \sigma$'s~\cite{DUNE:2015lol,DUNE:2026aaw,Hyper-Kamiokande:2025fci}. 



\section{Conclusions}\label{sec::Conclusions}
We have investigated the precision that can be achieved in the measurement of $\delta$ in long-baseline experiments, using DUNE as an example, assuming that all the runtime is spent studying neutrinos. In the standard configuration, the comparison between the oscillation probability in the neutrino and antineutrino sectors breaks the degeneracy with the other mixing parameters. On the other hand, however, the number of events in the antineutrino sector is considerably lower, leading to large statistical uncertainties. 

If only the neutrino sector is considered, the number of events is higher, but the degeneracies would reduce the sensitivity to $\delta$. The other mixing parameters, however, have already been measured accurately, and in the next few years the next generation of experiments will achieve sub-percent precision. If those constraints are taken into account, for instance by including penalty terms in the $\chi^2$, the sensitivity to $\delta$ increases significantly. The degeneracies can also be broken using the spectral information, {\it i.e.}, by examining the oscillation probability across different ranges of $L/E$. In accelerator experiments, this can be achieved, for example, by modifying the configuration of the magnetic horns, thereby selecting a different range of pion and kaon momenta and changing the neutrino beam energy. A precise estimation of the neutrino spectrum in such a case can be obtained through detailed simulations of the beam production. This, however, is beyond the purpose of this paper. For this reason, we used some approximations. First, we assumed that it is possible to shift the energy spectrum observed in the FHC mode to the right by $\Delta E$ without changing its shape. The other main assumption used regards the total flux luminosity. Indeed, one could naively expect that an increase of the neutrino energy would correspond to an increase of luminosity, due to the relativistic boost. However, a change in the configuration of the horns would likely affect the conversion efficiency as well. To see how much this factor could affect the sensitivity to $\delta$, we considered two cases: one in which the total flux normalization was left unchanged, ignoring the effect of the relativistic boost; another in which the flux was rescaled by a factor $(E_{peak}+\Delta E)^2/E_{peak}^2$. We have called it ``high-energy" configuration, while the case in which 100\% of the runtime is spent in neutrino mode, without changing the beam energy, was referred to as the ``only neutrino" configuration.

The precision that can be achieved in these set-ups was calculated with GLoBES simulations, using the AEDL files for the DUNE detector. This particular detector was chosen because, since it is on-axis, the neutrino energy spectrum seen by the far detector is broader than the one seen by off-axis detectors, and the second oscillation maximum can be observed as well. Moreover, changing the energy of the neutrino beam would be significantly more challenging if the detector is off-axis. We have found that, if the current or future uncertainties on the mixing parameter are taken into account, the ``only neutrino" configuration is outperforming the standard one for most of the values of $\delta$. If the relativistic boost is taken into account, the ``high-energy" configuration yields better results; the optimal choice for the energy shift is $\Delta E=1$ GeV. If the boost is not present, however, the ``only neutrino" configuration is slightly better. It is important to underline that the main goal of this paper was not to provide an exact estimation of the sensitivity to $\delta$ for a given experiment, but rather to investigate, qualitatively, if and under which conditions such an approach could offer some advantages. In particular, we want to point out that the results obtained studying the ``high-energy" configuration depend on the assumptions made: for a precise estimation of the sensitivity that can be achieved using this kind of set-up, detailed simulations of the neutrino beam production would be required. On the other hand, however, the results obtained using the ``only neutrino" configuration do not rely on such assumptions, since we used the spectra provided in the AEDL files, without modification.

Such an approach could increase the precision that can be achieved in future experiments, since its advantages would only increase as the precision of the measurements of the other mixing parameters improves, and this possibility should be taken into account in their designs.  

\section*{Acknowledgments}
I am deeply grateful to Jarah Evslin, Mauro Mezzetto, Maury Goodman, Lukas Berns and Bryce Littlejohn for the useful disccussions and comments.

\end{document}